\documentclass[mathleft]{an}
\usepackage[usenames,dvipsnames]{xcolor}
\usepackage{color}
\usepackage{graphicx}

\usepackage{times}

\def\solphys{Solar Phys.}
\def\apj{ApJ}
\usepackage{natbib}
\bibpunct{(}{)}{;}{a}{}{,}
\begin{document}

\Pagespan{1}{4}
\Yearpublication{2016}%
\Yearsubmission{2015}%
\Month{12}%
\Volume{999}%
\Issue{88}%

\title{Hybrid simulations of chromospheric HXR flare sources}

\author{Z. Moravec\inst{1}\fnmsep\thanks{Corresponding author:
  \email{zdenek.moravec@ujep.cz}\newline}, M. Varady\inst{1,2},
 J. Ka\v{s}parov\'{a}\inst{2}, \and D. Kramoli\v{s}\inst{1}}


\titlerunning{Hybrid simulations of chromospheric HXR flare sources}
\authorrunning{Z. Moravec et al.}
\institute{J.~E.~Purkyn\v{e} University, Faculty of Science, \v{C}esk\'{e}
              ml\'{a}de\v{z}e 8, CZ-400 96 \'{U}st\'{\i} nad Labem, Czech
              Republic
\and
Astronomical Institute of the CAS, Fri\v{c}ova 298, 25165 Ond\v{r}ejov, Czech Republic}

\received{14 Dec 2015}
\accepted{?????}
\publonline{later}


\abstract{%
Recent measurements of vertical extents and positions of the chromospheric hard
X-ray (HXR) flare sources based on Ramaty High-Energy Spectroscopic Imager
(RHESSI) observations show a significant
inconsistency with the theoretical predictions based on the
standard collisional thick target model (CTTM). Using a hybrid flare code
Flarix, we model simultaneously and self-consistently the propagation,
scattering and energy losses of electron beams with power-law energy
spectra and various initial pitch-angle distributions in a purely
collisional approximation and concurrently the dynamic response of the
heated chromosphere on timescales typical for RHESSI image reconstruction.
The results of the simulations are used to model the time evolution of
the vertical distribution of chromospheric HXR source within a
singular (compact) loop.  Adopting the typical RHESSI imaging times
scales, energy dependent vertical sizes and positions as could be
observed by RHESSI are presented.}

\maketitle

\section{Introduction}

RHESSI \citep{lin2002} observations of chromospheric
HXR sources in  solar flares present an invaluable diagnostics on the
properties of high energy electrons involved in the non-thermal
bremsstrahlung emission process in the thick target region
and on the properties of the target chromospheric plasma
\citep{brown1971}, providing thus an excellent basis for testing of
theoretical predictions of flare models.

The novel measurements of vertical sizes and positions of
chromospheric HXR sources in flares \citep[e.g.][]{batta_kont2011,
oliveros2012, krucker2015} based on the RHESSI imaging ability
\citep{lin2002, hurford2002} are one of such tests.
The source sizes in the direction of
the magnetic field measured by \citet{batta_kont2011} are in the range
from 1.3~arcsec up to 8~arcsec and hereby inconsistent with the
predictions given by the standard CTTM \citep{brown1971}. The
theoretical source
sizes obtained for static VAL~C atmosphere \citep{VAL1981} without
magnetic mirroring in the CTTM approximation are energy dependent,
but always smaller than the lower values obtained from observations,
for high energies even being sub-arcsecond
\citep[e.g.][]{moravec2013, varady2013}. The HXR source positions
for two studied footpoint sources according to \citet{oliveros2012} are
surprisingly close to the photosphere with heights from 200~km to
400~km and uncertainty $\pm200$~km, thus ruling out the CTTM.
\citet{krucker2015} obtained 
substantially higher positions for three flares, ranging from 600~km to 1050~km
including uncertainties. The upper values of these measurements are at
the lower limit predicted by CTTM. Their lower values, and especially those
obtained by \citet{oliveros2012}, rule out the standard CTTM and imply a
necessity of alternative models \citep{fletcher2008} or modifications
of the CTTM, e.g. by
introducing a re-acceleration of non-thermal electrons in the
chromosphere \citep{turkmani2006, brown2009, varady2014}.

There have been several attempts to explain the inconsistency between
the theoretical and observed vertical extents of the chromospheric HXR sources.
\citet{battaglia2012} modelled the vertical sizes of the HXR
sources for several prescribed density distributions along the loop,
for various pitch-angle distributions in a singular (compact) flare 
loop with converging magnetic field (i.e. including magnetic mirroring). They
obtained theoretical source sizes smaller than 1.5~arcsec, thus
inconsistent with the observed values. Another attempt \citep{flannagain2015}
attributes the discrepancy to the presence of neutral hydrogen
in the chromospheric thick target
region. The authors argue that the existence of
hydrogen atoms in the thick target region results in enhancement of
HXR emission in this region. Using their ad hoc atmospheres they were
able to extend the HXR source sizes to 3 arcsec on energies
between 40~keV and 60~keV.

\begin{figure}
\centerline{\includegraphics[width=40mm]{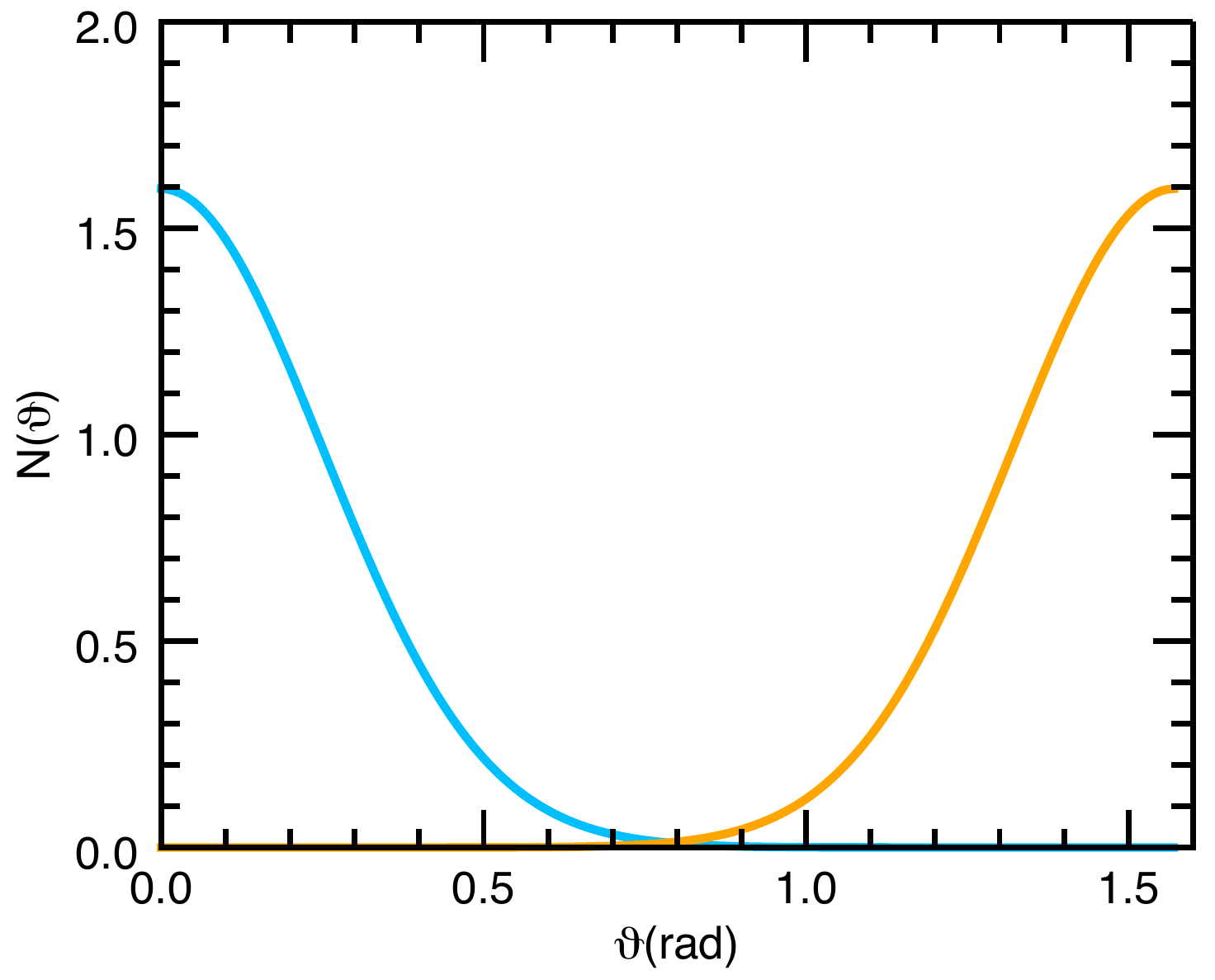}}
\caption{The pitch-angle distributions are gaussians with
  the standard deviation $\sigma=0.25$~rad. The light blue line corresponds to the strongly
  beamed distribution, the orange line to the pancake distribution.}
\label{fig:1}
\end{figure}

The motivation for this work is the following. Taking into account the
demands on sufficient photon statistics, the typical data needed for
quality RHESSI imaging are collected within several RHESSI rotations
corresponding to times $\sim10$~--~$10^2$~s (in \citet{batta_kont2011} it
was 60~s). Considering the
typical velocities of plasma flows in flare loops attained during the
first several tens seconds after the onset of the impulsive phase
(of orders $\sim10$ -- $100$~km~s$^{-1}$) obtained both
observationally from the Doppler shift measurements
in EUV/SXR \citep[e.g.][]{battaglia2015, milligan2006, milligan2009}
and from the hydrodynamic (HD) flare modelling \citep[e.g.][]{abbett1999,
  allred2005}, substantial changes in density and column density
distribution along the loop are inevitable and the time evolution
of the chromospheric HXR source can arise. The resulting size and position
of the HXR source detected  by RHESSI will be then
a result of superposition of HXR emitted in the flaring area during 
the whole accumulation time interval used for the image reconstruction.

In this paper we concentrate on theoretical modelling of vertical
sizes and positions of chromospheric HXR sources within a singular 
(compact) flare loop without magnetic field convergence.
The simulations are based on the  CTTM \citep{brown1971} in frame
of the standard CSHKP flare model \citep[see e.g.][]{shibata1996}. 
The model treats self-consistently the HD
response of the chromosphere to the energy released due to the Coulomb
collisions of non-thermal electrons with the chromospheric plasma on the
typical timescales corresponding to RHESSI photon collection times
needed for HXR image synthesis.

\begin{figure*}
\centerline{\includegraphics[width=158mm]{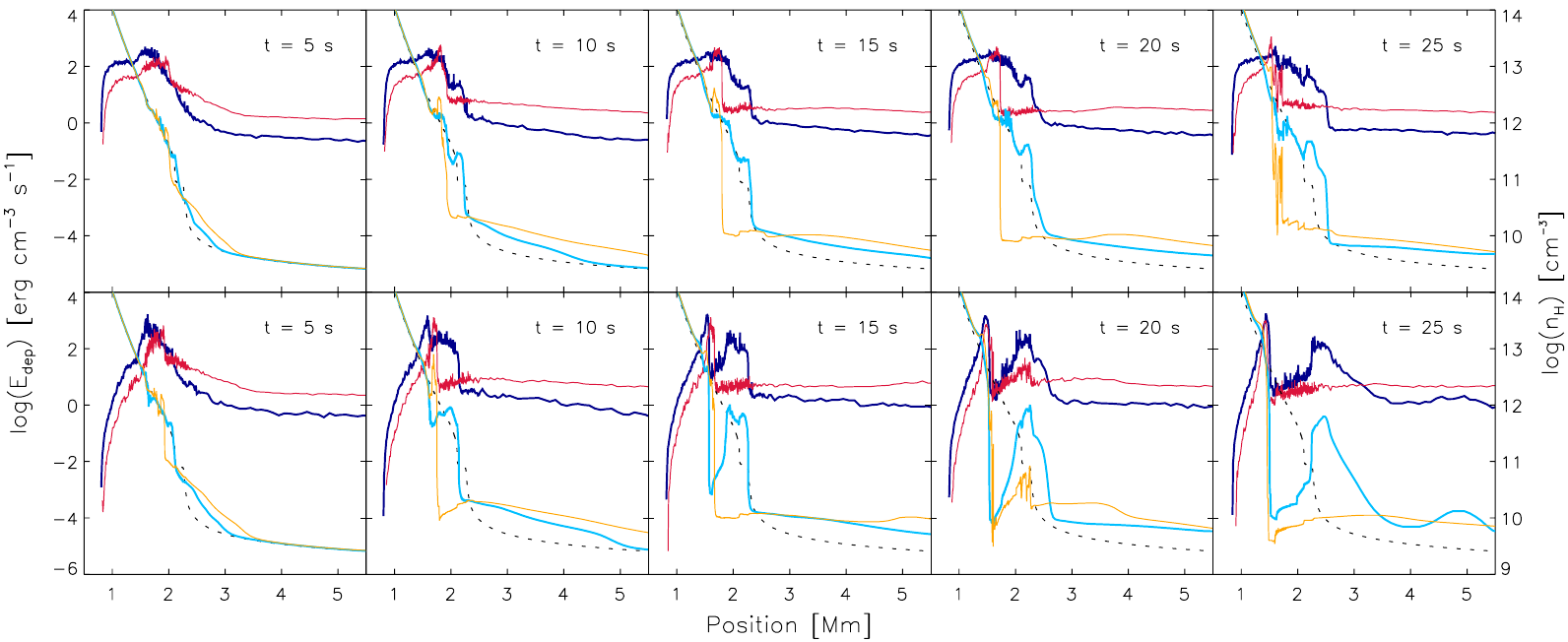}}
\caption{Time evolution of hydrogen number densities $n_\mathrm{H}$
  (light blue and orange lines) and energy deposits $E_\mathrm{dep}$
  (dark blue and dark red lines) for the first 25~s. The top panels correspond to $\delta=3$,
  the bottom panels to $\delta=7$. The thick blue lines denote the
  beamed pitch-angle distribution and flux
  $F_\mathrm{max}=2\times10^{10}$~erg~cm$^{-2}$~s$^{-1}$, the thin
  dark red and orange lines denote the pancake distribution and flux
  $F_\mathrm{max}=10^{10}$~erg~cm$^{-2}$~s$^{-1}$. The dotted black
  lines indicates the initial $n_\mathrm{H}$.}
\label{fig:2}
\end{figure*}

\section{Model description}

The simulations are performed using the hybrid code Flarix
\citep{kasparova2009, varady2010, varady2014} and the VAL C
atmosphere \citep{VAL1981} as the initial condition. We model
the atmospheric response to a single power-law electron beam generated
at the apex of a single flare loop of total length $L=15$~Mm and
a constant cross-section.

The physics related to the electron beam transport, scattering
and thermalisation due to Coulomb collisions \citep{emslie1978} is
modelled using a test-particle (TP) approach \citep{bai1982} based on
the Monte Carlo method. This approach, fully equivalent to the 
direct solution of the corresponding Fokker-Planck equation 
\citep{mackinnon1991}, provides a flexible
way to model many aspects of the non-thermal electrons interactions with
the ambient plasma, converging magnetic field along the loop or
with additional electric fields \citep{varady2014}.  Thanks to the 
TP approach, the detailed distribution function of the non-thermal
electrons is known at any instant time and position along the
loop. This information can be used to calculate
the distribution of HXR bremsstrahlung sources within the loop, their
positions, sizes, spectra and directivity of the emanating HXR 
emission (for details see \citet{varady2014}).

Concurrently with the transport, scattering and energy losses of the
non-thermal electrons, the HD response to the released
beam energy is calculated using a standard 1D
HD code for low-$\beta$ plasma.
It models the time evolution of the plasma within a semicircular
magnetic flux tube subjected to the heating by the electron beams.
Hydrogen ionisation is approximated as in \cite{brown1973}, the thermal conduction along
the field lines is treated using the standard Spitzer approximation \citep{spitzer1962}.
Total radiative losses consist of the optically thin losses 
from the transition region and corona \citep{rtv1978}
and an analytic approximation of the optically thick losses from the 
chromosphere \citep{peres1982}.
At the loop top in the corona we assume a plane symmetry, so only
  one half of the loop is modelled. The boundary conditions in the
  corona are a solid wall for mass, a reflecting one for momentum (or
  velocity) and zero derivative of temperature for the heat
  conduction. The backscattered TPs reaching the upper boundary
are reflected with a new pitch angle $\pi - \vartheta$ back
into the computational domain.

The 1D gas dynamics is treated
using the explicit LCPFCT solver \citep{oran2000}, the
Crank-Nicolson algorithm for the heat transfer and the time step
splitting technique to couple the individual source terms of the
energy equation with the HD. More details about the code can be found in
\cite{varady2010}.

\section{Results}

Here we discuss models of the electron beams of
the power-law indices $\delta =3, 7$, the maximum energy fluxes
$F_\mathrm{max}$,
and the low and high-energy cutoffs $E_0=20$~keV and $E_1=150$~keV, respectively.
Two types of spatial initial pitch-angle distributions are considered:
a strongly beamed and a pancake distribution (see~Fig.~\ref{fig:1}).
The energy fluxes are time dependent, linearly rising from $t=0$
to 2.5~s to their maximum values $F_\mathrm{max}$ and then being 
kept constant during the rest of the simulation, i.e. till $t=30$~s. The
simulations showed that the energy flux
$F_\mathrm{max}=2\times10^{10}$~erg~cm$^{-2}$~s$^{-1}$ 
gives too high temperatures in the corona  (over $10^8$~K) for the 
pancake distribution. The models for the pancake distribution are 
therefore calculated for reduced flux
$F_\mathrm{max}=10^{10}$~erg~cm$^{-2}$~s$^{-1}$, whereas the models
for the beamed distribution are calculated for
$F_\mathrm{max}=2\times10^{10}$~erg~cm$^{-2}$~s$^{-1}$.

\begin{figure}
\centerline{\includegraphics[width=75mm]{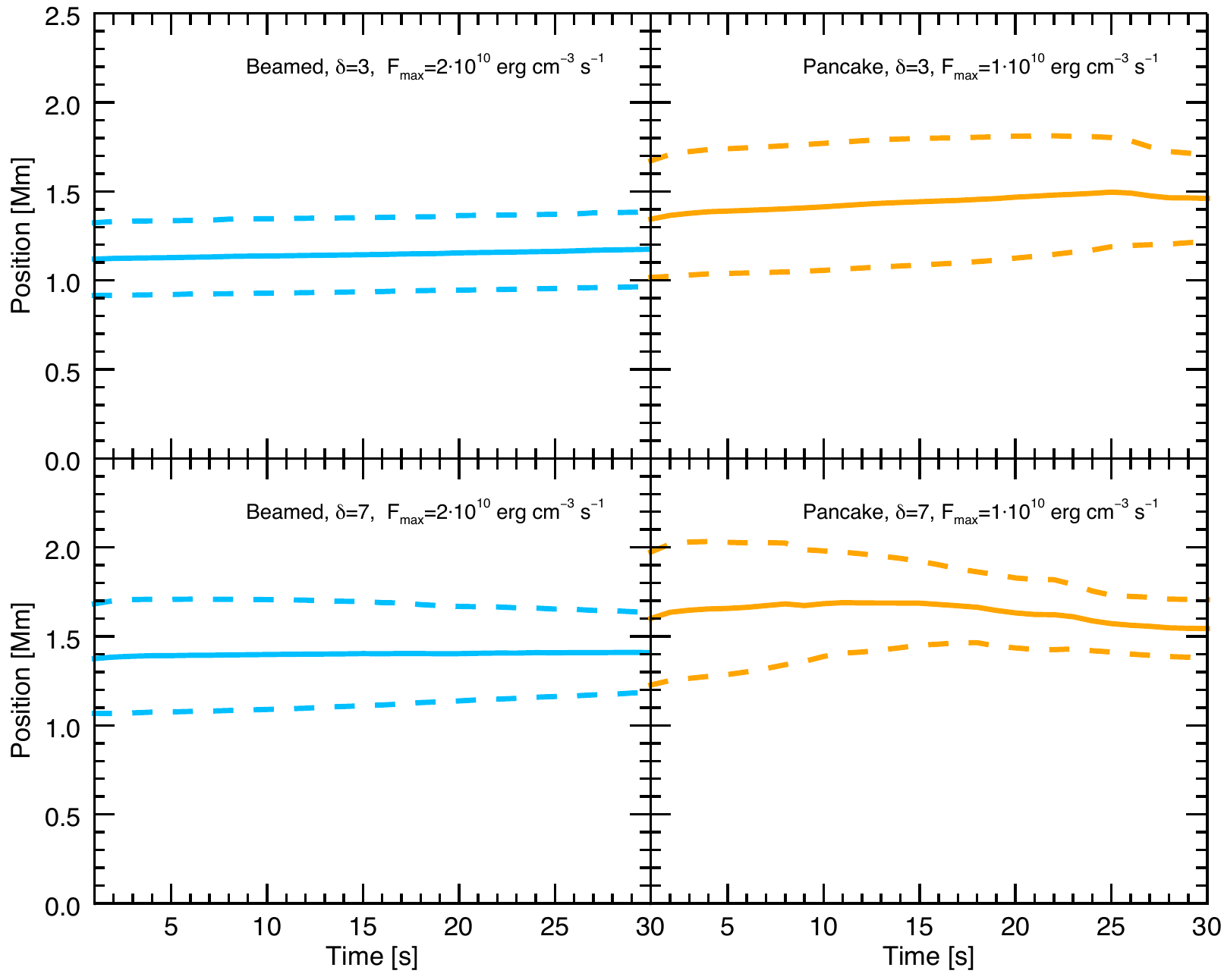}}
\caption{Time evolution of HXR chromospheric footpoint source positions (solid
  lines) and  FWHMs (dashed  lines) for integrated emission in
  30~--~70~keV range. The top panels correspond to
$\delta=3$, the bottom panels to $\delta=7$. The left panels show
the evolution for the beamed distribution and flux
$F_\mathrm{max}=2\times10^{10}$  (light blue), the right panels for the
pancake distribution and flux
$F_\mathrm{max}=10^{10}$~erg~cm$^{-2}$~s$^{-1}$ (orange).}
\label{fig:3}
\end{figure}

\begin{figure}[!h]
\centerline{\includegraphics[width=75mm]{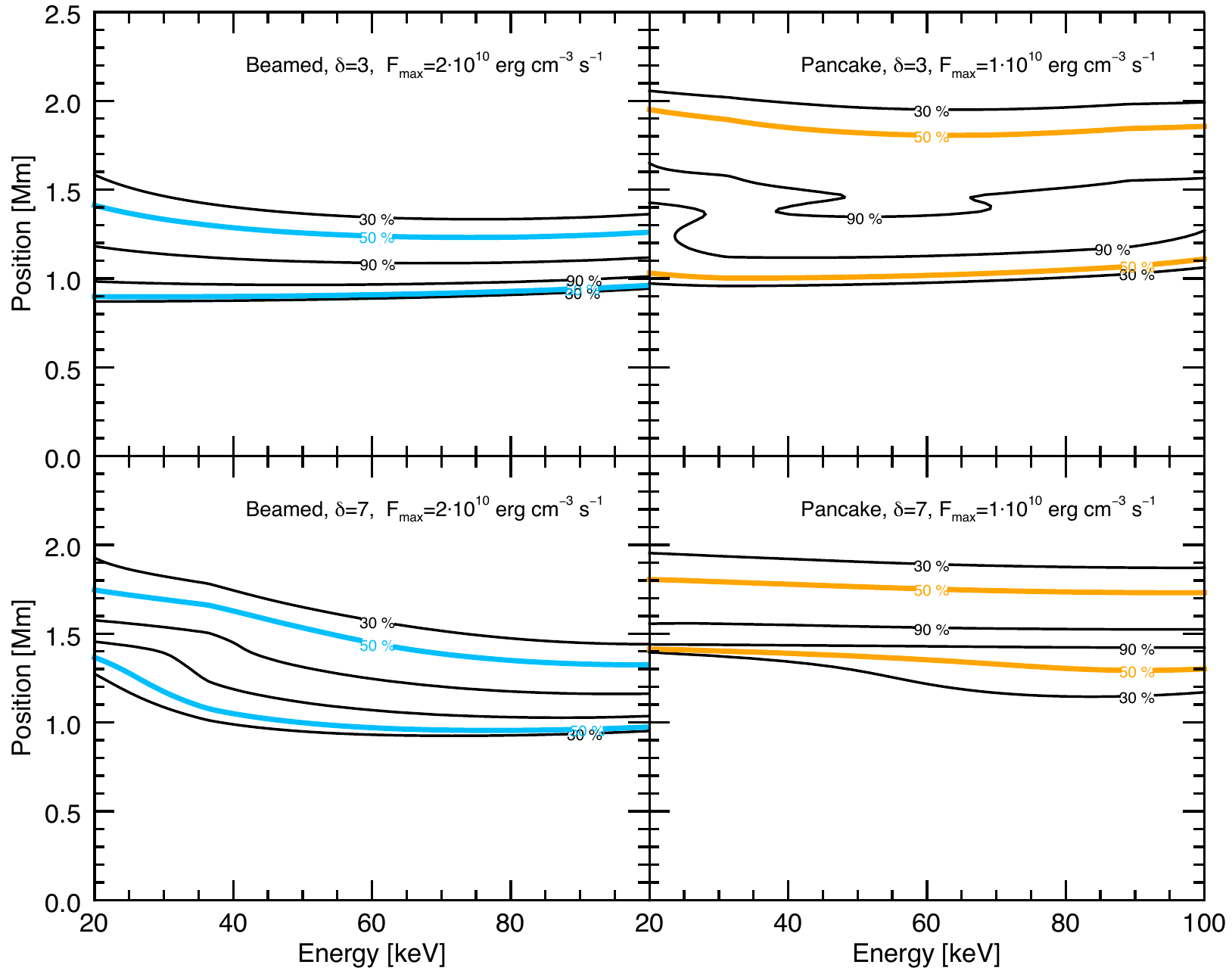}}
\caption{ Chromospheric footpoint HXR sources accumulated during the whole 30~s simulation,
as a function of energy and height. The positions of individual models in the panels
and color coding correspond to Fig.~\ref{fig:3}.}
\label{fig:4}
\end{figure}

The time evolutions of the hydrogen number densities and
energy deposits along the beam heated atmosphere 
are shown in
Fig.~\ref{fig:2}. The patterns in the density evolution significantly
vary model from model. The 'pancake' electrons pass, thank to their
pitch angles, much greater column densities per unit distance along
the loop axis than the 'beamed' electrons. That is why they tend to
deposit their energies in low density, upper layers of the
atmosphere. This results in strong heating and evaporation without
formation of any significant density waves. The effect is obvious for
both $\delta$, but naturally, it is more pronounced for $\delta=7$.

The 'beamed' electrons tend to reach deeper and denser layers, where
the radiative losses drain a substantial part of the deposited energy,
reducing thus the energy that can be consumed on heating and evaporation.
The importance of this effect increases with spectral hardness
(compare the light blue lines for $n_\mathrm{H}$ in the top and bottom panels
of Fig.~\ref{fig:2}) and results in lower plasma densities in the
corona relative to the pancake distribution, at least for times
$\le20$~s. Probably because the 'beamed' energy deposits
  reach deeper and denser layers they tend to form strong density
  waves propagating with high velocities into the corona
  (see~Fig.~\ref{fig:2}). The time evolution of the energy deposits
  corresponds to the density structure of the atmosphere.

To describe the characteristics of HXR emission in the modelled
loop, we follow the approach of \cite{battaglia2012} and define the
position of the source as the first moment of the height profile of
HXR emission and the FWHM as the second moment of that profile.
Similarly, only emission above 10\% of the maximum HXR is considered
in order to emulate the limited dynamic range of RHESSI images.

Although the density and temperature structure significantly
changes (see~Fig.~\ref{fig:2}), the positions and FWHMs of the HXR
sources  integrated in the 30~--~70~keV range change only modestly
(see~Fig.~\ref{fig:3}). This applies especially to the models
with the beamed distribution, where the sources are
static at position $1.1$~Mm with FWHM $0.6$~arcsec  for
$\delta=3$ and $1.4$~Mm with FWHM from 0.9 to 0.6~arcsec for
$\delta=7$ (see left panels in Fig.~\ref{fig:3}).
For the pancake distribution and $\delta=3$ we obtained the
largest HXR source with FWHM 1.2~arcsec, placed at a constant height
1.4~Mm (see the top right panel in Fig.~\ref{fig:3}). Also the source for
$\delta=7$ (see the bottom right panel in Fig.~\ref{fig:3}) remains at
constant position around 1.6~Mm, but its FWHM decreases with time
from 1.1~arcsec to 0.4~arcsec. Generally, the sources
corresponding to models with $\delta=3$ are located lower than
those for $\delta=7$, and the sources for pancake distribution are
located higher than those for the beamed distribution 
(Figs.~\ref{fig:3}, \ref{fig:4}).

Fig.~\ref{fig:4} shows the energy and height structure of HXR sources
integrated through the duration of 30~s heating. Models of beamed
pitch-angle distribution result in lower and smaller HXR sources,
whereas pancake distributions correspond to larger sources located
higher in the atmosphere. Both characteristics of the HXR height profile
are almost  independent on energy, in the 20~--~100~keV range.

\section{Conclusions}

Using the models of HD response to the beam heating, we 
studied characteristics of chromospheric flare HXR sources
for the first 30~s after the start of the energy deposit for various
parameters of the electron beam and two very distinct initial 
pitch-angle distributions.
\begin{enumerate}
\item In all models the assumed beam heating resulted in significant
changes of density and temperature structure of the flare loop.
\item Despite the substaintial change of the atmosphere structure, the
position and FWHM of the HXR sources in the 30~--~70~keV range show
very weak time dependence. This is true especially for the beamed
pitch-angle distribution. The model of $\delta=7$ and pancake
distribution indicates a change in FWHM with time.
\item Time averaged HXR emission through 30~s beam heating is
not significantly dependent on the energy. HXR sources of the beamed
distribution are smaller and located lower in comparison with models
of pancake distribution.
\item HXR source characterictics are weakly related to the values of
  energy flux or beam power-law index, however, they show dependence
  on pich-angle distributions, at least for two extreme cases studied
  in this paper.
\item All studied models give much smaller HXR sources than observed
ones, this is in agreement with previous parametric study of
\citep{battaglia2012}. The largest HXR source, 1.2~arcsec, is found
for $\delta=3$ and pancake distribution, but its location is higher
than those from observations \citep{krucker2015} .
\item Next steps are to study influence of converging magnetic field
   and the effect of multi-thread structure.
\end{enumerate}

\acknowledgements
 We acknowledge funding from the European Community's Seventh 
Frame Programme (FP7/2007-2013) under grant agreement 
no.~606862 (F-CHROMA). D. Kramoli\v{s} and M. Varady
also acknowledge the support of the UJEP Student Grant
Agency. The calculations were performed on Enputron, a computer
cluster at the J. E. Purkyn\v e University,
Faculty of Science.

\bibliographystyle{aa}
\bibliography{potsdam_1.0_ref}

\end{document}